\documentclass[ag]{copernicus}

\usepackage{mathptmx}

\usepackage{graphicx} 

\usepackage{amsmath}
\usepackage{amsfonts}
\usepackage{amssymb}
\usepackage{amsbsy}
\usepackage[figuresright]{rotating}
\usepackage{multicol}
\usepackage{array}
\usepackage{supertabular}

\usepackage{color}

\begin{document}

\title{{A note on the Weibel instability  and thermal fluctuations}}

\author[1,2]{{R. A. Treumann}
}
\author[3]{{W. Baumjohann}}

\affil[1]{Department of Geophysics and Environmental Sciences, Munich University, Munich, Germany}
\affil[2]{Department of Physics and Astronomy, Dartmouth College, Hanover NH 03755, USA}
\affil[3]{Space Research Institute, Austrian Academy of Sciences, Graz, Austria}

\runningtitle{Weibel instability and thermal fluctuations}

\runningauthor{R. A. Treumann and W. Baumjohann}

\correspondence{R. A.Treumann\\ (rudolf.treumann@geophysik.uni-muenchen.de)}

\received{ }
\revised{ }
\accepted{ }
\published{ }


\firstpage{1}

\maketitle

\begin{abstract}
The thermal fluctuation level of the Weibel instability is recalculated. It is shown that the divergence of the fluctuations at long wavelengths, i.e. the Weibel infrared catastrophe, never occurs. At large wavelengths the thermal fluctuation level is terminated by the presence of even the smallest available stable thermal anisotropy. Weibel fields penetrate only one skin depth into the plasma. When excited inside, they cause layers of antiparallel fields of skin depth width and vortices which may be subject to reconnection.

\end{abstract}

\introduction
\vspace{3mm}\noindent
The Weibel thermal \citep{weibel1959} and Weibel filamentation \citep{fried1959} instabilities, two closely  related versions of the purely electromagnetic zero frequency instability of a non-magnetised plasma, have been proposed \cite[see, e.g.,][]{yoon1987} as fundamental for the generation of quasi-stationary magnetic fields \citep{medvedev1999} in high temperature or high-speed  \citep[cf., e.g.,][]{achterberg2007} plasmas, respectively. The mechanism of this instability has been elucidated \citep{fried1959} as the spontaneous  microscopic electron (or ion) currents flowing in a plasma which, under unstable conditions, become amplified and cause a non-vanishing magnetic field to grow in the plasma. As usual, the field grows up to a maximum amplitude and stabilises by some mechanism like exhaust of the available free energy, in the case of the thermal Weibel mode quasilinear depletion of the temperature anisotropy {\citep[recent particularly complete discussions are given in][]{pokh2010,pokh2011}}, in the case of the Weibel filamentation mode by self-magnetisation of the initially non-magnetised streaming particles \citep[cf., e.g.,][]{achterberg2007}. The generated  fields are (quasi)-stationary on temporal scales longer than growth and stabilisation times. {Nonlinear effects set on when the waves reach large amplitudes after amplification. At this stage, quasilinear modification of the distribution function limits further growth \citep{pokh2011} and wave-wave interaction causes further nonlinear evolution \citep{pokh2010} which distributes particle energy into other modes, further limiting the growth of magnetic fields. Nevertheless, under non-dynamo conditions, t}he Weibel mechanism is thus an alternative to the celebrated turbulent magnetic dynamo mechanism, the latter being believed to be responsible for the generation of magnetic fields in planets \citep[][]{christensen2010,hulot2010}, stars \citep[e.g.][]{jones2010}, galaxies and on the large scale in the universe. The Weibel instability may grow in the universe from thermal fluctuations at early times \citep{widrow2011} either in the chromodynamic state or later in the plasma state, and also after recombination during structure formation \citep{ryu2011}. It is believed to dominate all other mechanisms in magnetic field generation in  relativistic shocks \citep[cf., e.g.,][]{byk2011}. More recently, it has been suggested to play a decisive role in the ignition of magnetic reconnection in collisionless current sheets \citep{baum2010,treumann2010}. These facts, in particular the astrophysical application to magnetic field generation in shocks, let the Weibel instability attract considerable interest.  Nevertheless, several open questions remain in dealing with it. These refer to the instability itself, independent of its free energy source, i.e. whether thermal or flow driven, and to its thermal level. In the following we very briefly clarify two of these: the nature of the Weibel mode, and avoiding the infrared catastrophe of Weibel thermal fluctuations.

\section{Weibel mode}
In which mode does the Weibel instability propagate? This question is non-trivial because, in unmagnetised plasma of density $N$, electromagnetic wave propagation becomes possible only at frequencies above the plasma frequency $\omega_e=\sqrt{e^2N/\epsilon_0m_e}$, a condition which immediately follows from the unmagnetised plasma wave dispersion relation $\omega^2=k^2c^2+\omega_e^2$, with $k$ wavenumber, and $c$ velocity of light. Zero frequency modes cannot propagate in this case. As a by-product this implies that no magnetic fields can be imposed into the collisionless plasma. They must have been present from the very beginning, i.e. they must have been born with the plasma. Otherwise magnetic fields can enter only a distance of the order of the electron skin (inertial) depth, $\lambda_e=c/\omega_e$ in the collisionless case. Such waves in the plasma are evanescent.  As is well known, the situation changes, when the plasma is already magnetic, containing a constant magnetic field $\mathbf{B}$, in which case Alfv\'en, whistler, and magnetosonic waves can propagate down to extremely low frequency. Hence, what is the reason for the Weibel instability to survive and cause real magnetic fields in plasma?
\begin{figure}[t!]
\centerline{{\includegraphics[width=0.3\textwidth,clip=]{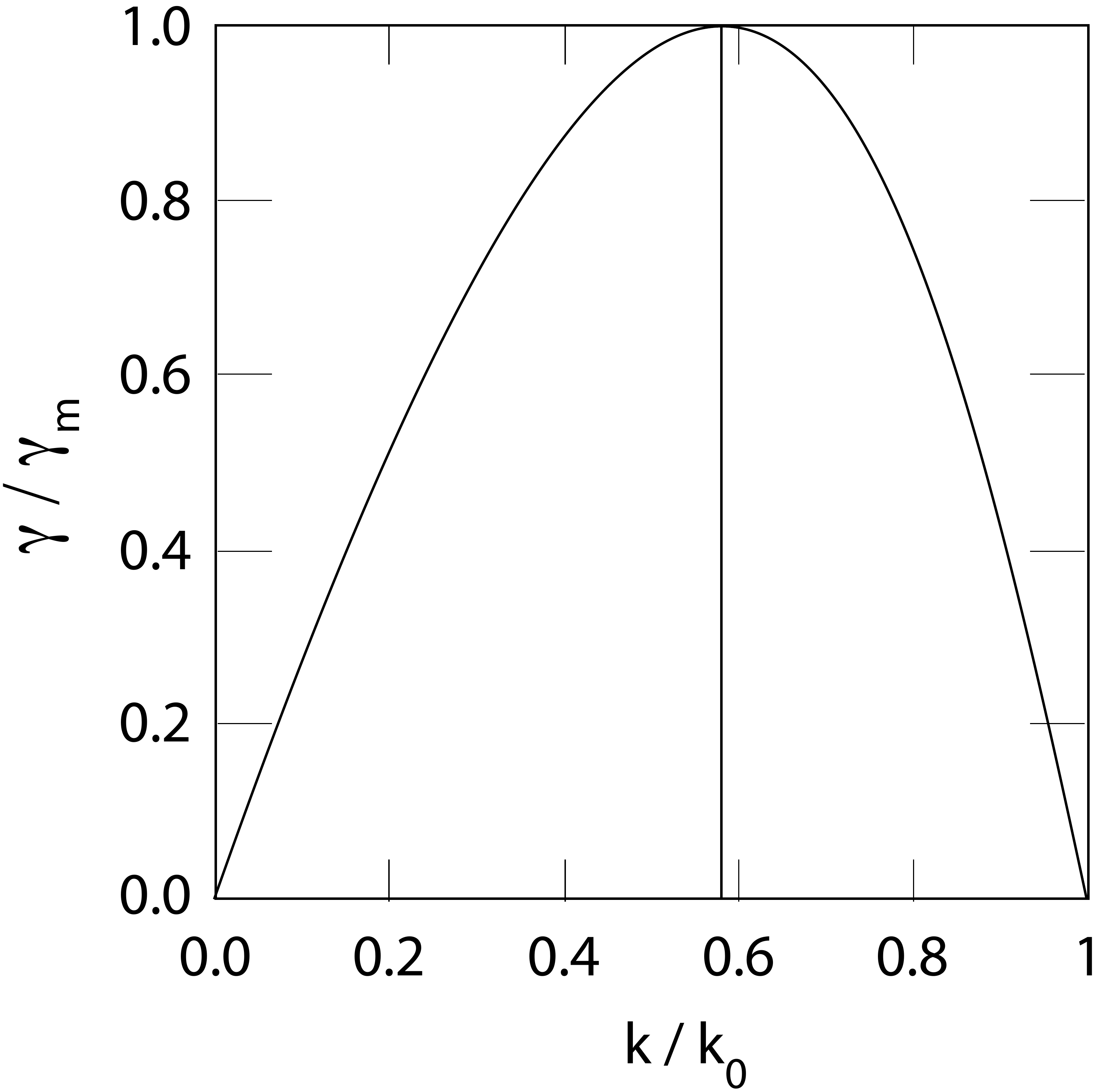}
}}\vspace{-3mm}
\caption[ ]
{\footnotesize {Dependence of the normalised (to maximum growth $\gamma_m$) Weibel growth rate $\gamma$ on wavenumber. The domain of growing wave numbers is limited by the skin-dpeth wave number $k_e=\omega_e/c$.}}
\label{fig-weib-one}
\end{figure}

\begin{figure}[t!]
\centerline{{\includegraphics[width=0.4\textwidth,clip=]{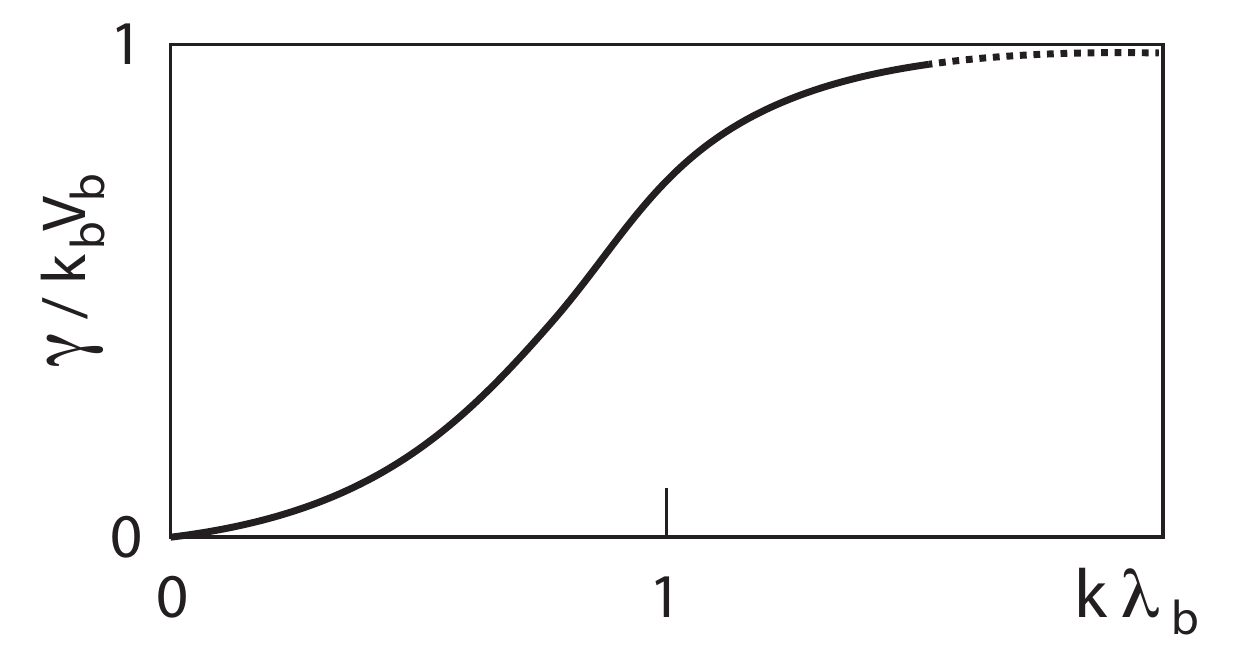}
}}\vspace{-3mm}
\caption[ ]
{\footnotesize {The growth rate of the Weibel filamentation instability as function of wavenumber $k\lambda_b\equiv k\lambda_{eb}\sqrt{N_b/2N_e}$.}}
\label{fig-weib-fil}
\end{figure}

In order to understand this point, we briefly repeat the thermal Weibel theory giving it a simple representation.  Thermal Weibel instability requires that the plasma temperature is -- for some unspecified reason -- anisotropic, allowing to define two temperatures $T_\|, T_\perp$ in orthogonal directions. For instance, if the plasma streams with velocity $\mathbf{V}$, then the plasma pressure might become anisotropic,
\begin{equation}
\textsf{P}_e=nk_B\Big[T_{e\perp}\textsf{I}+\big(T_{e\|}-T_{e\perp}\big)\mathbf{VV}/V^2\Big]
\end{equation}
(with $\textsf{I}\equiv \delta_{ij}$, $(i,j)=1,2,3$ the unit tensor) thus defining anisotropic temperatures parallel and perpendicular to the flow, yielding a bi-Maxwellian equilibrium distribution. The free energy stored in the anisotropy is available for driving instability. The linear dispersion relation \citep[see, e.g.][]{treumann1996}, from which instability can be inferred, becomes
\begin{equation}
D(\omega,\mathbf{k})=\left(n^2-\epsilon_\perp\right)^2\epsilon_\|=0, \qquad n^2=k^2c^2/\omega^2
\end{equation}
where $\epsilon_\perp(\omega,\mathbf{k}), \epsilon_\|(\omega,\mathbf{k})$ are the transverse and longitudinal response functions. The longitudinal function is of no interest as it does not yield electromagnetic waves. The transverse response function is, using the bi-Maxwellian, given by 
\begin{equation}
\epsilon_\perp(\omega,\mathbf{k})= 1+\frac{\omega_e^2}{\omega^2}\Big\{1-\big(A+1\big)\Big[1+\zeta \,Z\big(\zeta\big)\Big]\Big\} -\frac{\omega_i^2}{\omega^2}
\end{equation}
Here $Z(\zeta)$ is the plasma dispersion function of argument $\zeta=\omega/k_\perp v_{e\perp}$, and $A=T_{e\|}/T_{e\perp}-1>0$ the thermal anisotropy, $v_e^2=2T_e/m_e$ the thermal velocity, and $\mathbf{k}=(k\sin\theta, 0,k\cos\theta)$ is the wave vector having only two components. At very low frequencies $\omega\approx 0$ the unstable solution obtained by putting $\omega(\mathbf{k})=i\gamma(\mathbf{k})$ purely imaginary, yields for the Weibel growth rate
\begin{equation}
\gamma_{\,WI} (k_\perp)=\sqrt{\frac{2}{\pi}} \frac{k_\perp v_{e\perp}}{k_0\lambda_e}\left(1-\frac{k_\perp^2}{k_0^2}\right)\big(A+1\big)\big(k_0\lambda_e\big)^3
\end{equation}
With $A>0$, the wave will grow as long as $k_\perp<k_0$, where $k_0\lambda_e=\sqrt{A}$ limits the unstable range in wavenumber. Maximum growth is obtained at $k_m=k_0/\sqrt{3}$, close to the instability cut-off, This is shown in Fig. \ref{fig-weib-one} with maximum growth rate 
\begin{equation}
\gamma_m(k_m)= \omega_e\sqrt{\frac{8\, A^3}{27\pi}} \left(\frac{v_{e\perp}}{c}\right)(A+1)\ll 
\omega_e
\end{equation}
thus justifying the initial assumption of small frequency,  
but the unstable waves being of wavelength somewhat larger than the  plasma skin depth. We note in passing that the Weibel filamentation growth rate for two equal density counter streaming electron flows of speed $V_b$ becomes
\begin{equation}
\gamma_{\,FI}= k_{be}V_b\Bigg\{1+\frac{\omega_{be}^2}{k^2c^2}\Bigg[1+\frac{2N_e}{N_b}\Big(1+\frac{m_e}{m_i}\Big)^{-1}\Bigg]\Bigg\}^{-\frac{1}{2}}
\end{equation}
where the index $b$ refers to the two streaming plasmas. In the long-wavelength range, if the background density is large, $N_e>N_b/2$, this expression simplifies to 
\begin{equation}
\frac{\gamma_{\,FI}}{k_{be}V_b}\lesssim  k\lambda_{eb}\sqrt{\frac{N_b}{2N_e}}\Big(1-\frac{N_b}{4N_e}\Big)\equiv k\lambda_b
\end{equation}
where $k_{be}=\omega_{be}/c=\lambda_{be}^{-1}$. This growth rate is small, unless $V_b$ is large. The plot of the growth rate is given in Fig. \ref{fig-weib-fil}. It shows, in contrast to the thermal Weibel mode, that the linear growth rate is not restricted in the first place but formally extends to the limit $\gamma_{\,FI}/k_{be}V_b=k\lambda_b$ for $k\to\infty$. Of course, a natural upper bound of $k<k_D$ is provided by the Debye wave number $k_D$.

From this one concludes that, in a strictly non-magnetised plasma, the Weibel mode, whether thermal or flow driven, is not a bulk plasma mode. Under unstable conditions when it is excited in the plasma at some location $x$, its magnetic field will be felt only in an interval  $[x-\lambda_e,x+\lambda_e]$ around $x$, producing magnetic fields in a region of the size of $\sim 2\lambda_e$. Therefore, in an extended plasma containing temperature anisotropies, the Weibel mode will necessarily cause filamentary magnetic field structures, even when excited by thermal anisotropies. These structures, because of the zero divergence requirement are located adjacently carrying Weibel magnetic fields of opposite direction on transverse scales of the electron skin depth. Such structures resemble current layers, may be unstable against reconnection, form magnetic islands and dissipate part of the free energy by transforming it into plasma heat, depletion of temperature anisotropy or kinetic flow energy, and related effects.

\section{Thermal fluctuation effects}
The Weibel instability is not really of zero-frequency.  
An estimate of the finite though even smaller real frequency $\omega_r$ is obtained from the real part of the dispersion relation (formally assuming $\gamma\gg\omega_r$)  under the simplifying assumption that $\omega \ll \omega_e$. This yields
\begin{equation}
\omega_r\approx k_\perp c\left(\frac{v_{e\perp}}{c}\right)\sqrt{\frac{1}{2}\frac{A+m_e/m_i}{A+1}}
\end{equation}
Since $v_e/c\ll 1$ this shows that the frequency is very small, the index of refraction is $n>1$, and the wave, if propagating at all,  must propagate on one of the low frequency magnetised plasma branches, Alfv\'en or whistler. The long wavelength whistler dispersion relation is
\begin{equation}
\frac{k^2c^2}{\omega^2}\approx\frac{\omega_e^2}{\omega\omega_{ce}} 
\end{equation}
with $\omega_{ce}=eB/m_e$ the electron cyclotron frequency. In order for the Weibel mode to propagate at such low frequency as a whistler (or ultimately electron Alfv\'en) wave, this requires that the plasma must be weakly magnetised initially and
\begin{equation}
k_\perp V_{Ae}/\omega_e\approx \sqrt{A/(A+1)}
\end{equation}
determines the wave number, with $V_{Ae}$ the electron Alfv\'en speed. This becomes possible once the Weibel instability has grown in the above spatial slot $[x-\lambda_e,x+\lambda_e]$. 

However, initially, the plasma is not completely free of magnetic fields. Thermal fluctuations provide a weak magnetic fluctuation background level. This has been known for long time \citep{landau1959,landau1960} for any thermal electrodynamically active medium. The explicit theory of thermal fluctuations in plasmas can be found in \citet{sitenko1967} and \citet{akh1975}. Application to magnetic fluctuations has been given by \citet{yoon2007} and \citet{treumann2010}. Being a thermal dielectric response property, thermal fluctuations can exist in regions, where real waves do not propagate. Otherwise, they provide a pool of initial rms wave amplitudes, $\bar b=\sqrt{\langle b^2\rangle}$, from which the instability chooses to excite a particular range of wave modes which are allowed to propagate. For the Weibel mode this property has already been exploited \citep{treumann2010}. 

On the other hand, the weak magnetic fluctuation level, $\bar{b}$, also provides an average finite weak background magnetic field which can be chosen by the unstable wave to propagate. Taking this fluctuation wave field, the background Alfv\'en velocity becomes $\langle V_A\rangle=\bar{b}/\sqrt{\mu_0 m_eN_e}$, which enters the above whistler dispersion relation and fixes the wave number of the unstable modes which then may, indeed, propagate. 

Since the frequency becomes very low, implying that also $k_\perp\to 0$, when interested in the generation of quasi-stationary magnetic fields, the magnetic fluctuation spectrum must be calculated slightly more precisely than done in the above references. There the magnetic fluctuation level depended on wave number as $\langle b^2\rangle\propto (k_\perp\lambda_e)^{-3}$, leading to an `infrared catastrophe' for vanishing $k_\perp$ and quasi-stationary fields. 

Such a catastrophe cannot occur in nature. A more precise calculation should take into account two facts. First, the plasma also contains ions (which have been included in the above transverse response functions but have not yet been used). Second, calculation of thermal fluctuations in the thermodynamic equilibrium (or quasi-equilibrium) state for  the stable thermal Weibel mode implies  
\begin{equation}
A=T_\|/T_\perp-1 <0
\end{equation}
The transverse response function can be written as
\begin{equation}
\epsilon_\perp=1+\frac{\omega_e^2}{\omega^2}\Bigg\{1-\frac{T_\|}{T_\perp}\Big[1-\Phi(z)+i\sqrt{\pi}\,z\,\mathrm{e}^{-z^2}\Big]\Bigg\}-\frac{\omega_i^2}{\omega2}
\end{equation}
where $z=\zeta/\sqrt{2}$, and $\Phi(z)\approx 2z^2$ for $z\ll 1$. Then, in the very low frequency limit, $\omega\to0$, one obtains the correct thermal spectral energy density of magnetic fluctuations in the  anisotropic-pressure plasma as
\begin{equation}
\langle b^2(k)\rangle=\frac{\mu_0}{\omega_e}\sqrt{\frac{\pi T_\perp}{m_ec^2}}\frac{m_ec^2(A+1)^2k\lambda_e}{(A+2)\big[k^2\lambda_e^2-A-m_e/m_i\big]^2}
\end{equation}

For $A>0$ we had $k^2\lambda_e^2\leq A$, avoiding catastrophe and providing a thermal background to chose for the Weibel instability. For $A<0$, the stable case, infrared catastrophe is inhibited by the presence of ions until $m_e/m_i<-A$ and, with $k\to0$, for $-2<A$.  However, when $A\leq -2$, the nature of the fluctuations changes. In this case magnetic fluctuations belong to the firehose range which excludes the Weibel mode anyway.

The above spectral energy density is a function of wavenumber $k$. It can, in principle, be used in order to determine the wavelength range in which the Weibel mode propagates on the lowest frequency whistler (or Alfv\'en) branch.  This requires calculation of the rms thermal fluctuation amplitude and inserting it into the Alfv\'en velocity. We do not perform this simple step here.

\conclusions
The intent of this note, which we believe having achieved, was a critical examination of the role of the Weibel instability. The Weibel instability, though known for half a century, has only recently been celebrated  as a new and apparently important non-dynamo generator of magnetic fields in high temperature plasmas. By theory, it would provide relatively long-scale, $0\lesssim k_\perp\lambda_e<1$, quasi-stationary magnetic fields. The linear growth rate, however, maximises in the short-scale range, close to the electron skin depth (or ion skin depth in the ion equivalent of the Weibel mode, if excited, for instance in high speed ultra-relativistic flows). 

We have suggested that another restriction limits the wavelength, i.e. the inhibition of electromagnetic modes to propagate in dense non-magnetised plasmas. On the other hand, thermal fluctuations in such plasmas allow for a low level of rms magnetic field amplitudes which serve as background magnetisation. These will always be present, thereby opening a channel for the Weibel mode to propagate on the lowest frequency Alfv\'en or whistler mode branches. 

At these (quasi-stationary) frequencies, thermal fluctuations were believed to undergo an infrared catastrophe. This is not the case. Taking into account the ion contribution  and respecting the further condition that thermal fluctuations evolve in the range of stable plasma states, fluctuation theory is valid up to infinitely long wavelengths. The magnetic spectral energy density behaves analytically. 

Though the weak rms magnetic fluctuation background indeed solves the problem of Weibel penetration into a plasma, the Weibel magnetic fields are on scales not much longer than the skin depth. Divergence conditions then require that these fields form vortices of neighbouring antiparallel fields, resembling current layers. These may undergo reconnection, forming islands and thus limit the Weibel amplitude growth in various ways: by structuring the field, plasma jetting, and heating the plasma until depletion of the anisotropic temperature, which is the free energy source of the instability.

\begin{acknowledgements}
This research was part of an occasional Visiting Scientist Programme in 2006/2007 at ISSI, Bern. 
\end{acknowledgements}

\end{document}